# Determination of the calcium species in coal chars by Ca K-edge XANES analysis[*]

LIU Lijuan(刘利娟)[1], LIU Huijun(刘慧君)[2], CUI Mingqi(崔明启)[1,1)], HU Yongfeng(胡永峰)[3], ZHENG Lei(郑雷)[1], ZHAO Yidong(赵屹东)[1], MA Chenyan(马陈燕)[1], XI Shibo(席识博)[1], YANG Dongliang(杨栋亮)[1], GUO Zhiying(郭志英)[1], WANG Jie (王杰)[2,2)] ,

[1] Institute of High Energy Physics, Chinese Academy of Sciences, Beijing 100049, P. R. China

[2] Department of Chemical Engineering for Energy, East China University of Science and Technology, Shanghai 200237, China

[3] Canadian Light Source, University of Saskatchewan, Saskatoon, Saskatchewan, S7N 0X4, Canada

**Abstract:** Ca-based additives have been widely used as a sulfur adsorbent during coal pyrolysis and gasification. The Ca speciation and evolution during the pyrolysis of coal with Ca additives have attracted great attention. In this paper, Ca species in the coal chars prepared from the pyrolysis of $Ca(OH)_2$ or $CaCO_3$-added coals are studied by using Ca K-edge X-ray absorption near edge structure spectroscopy. The results demonstrate that $Ca(OH)_2$, $CaSO_4$, CaS and CaO coexist in the $Ca(OH)_2$-added chars, while $Ca(OH)_2$ and $CaSO_4$ are the main species in the $Ca(OH)_2$-added chars. Besides, a carboxyl-bound Ca is also formed during both the pyrolysis for the $Ca(OH)_2$-added and the $CaCO_3$-added coals. A detailed discussion about the Ca speciation is given.

## 1 Introduction

Sulfur in coal is generally unfavorable for the utilization of coal, because it forms the environmental pollutant. Capture of sulfur-containing gases during coal utilization was a common way to reduce the emissions of sulfur-containing gases. Ca-based additives, such as CaO, $CaCO_3$ and $Ca(OH)_2$ have been widely used as a sulfur adsorbent during coal pyrolysis and gasification [1-5]. It is well known that the efficiency of sulfur retention differs with Ca-based additives. X-ray diffraction (XRD) confirmed that $Ca(OH)_2$ had a higher efficiency of capturing sulfur than $CaCO_3$ [2]. Although XRD analysis is widely used to determine the Ca-based minerals in raw coal and chars, it is difficult and even unavailable to determine the Ca species with non-crystalline structure including the calcium associated with the organic ligands. X-ray absorption near-edge structure (XANES) spectroscopy is an element-selective and local-structure sensitive technique, and it can be used to identify the variety of calcium species in coal and other materials.

There have been some Ca K-edge XANES studies. For example, Ca K-edge XANES was used to qualitatively and quantitatively determine Ca speciation [6-12]. A few studies focused on the changes of Ca species during coal pyrolysis, gasification [6,9,11,12]. Besides the Ca speciation, the percentage composition of different Ca species in raw coal was also obtained by means of Ca K-edge XANES [10]. Although $CaCO_3$ and $Ca(OH)_2$ have been often used as Ca-based additives to capture sulfur during pyrolysis of high-sulfur coal, the information regarding the transformation of Ca species is still unavailable. Especially, the change to non-crystalline calcium species in the pyrolysis process was scarcely reported. In this study, Ca K-edge XANES is used to determine the Ca species in chars prepared from $CaCO_3$–added and $Ca(OH)_2$–added coals.

[*] Supported by National Program on Key Basic Research Project (2010CB227005-03) and National Natural Science Foundation of China (21076081 and 10775150)
[1)] E-mail: cuimq@ihep.ac.cn
[2)] E-mail: jwang2006@ecust.edu.cn

## 2 Experiment

The samples used in this study were a high-sulfur bituminous coal from Chongqing city in China. The chemical analysis for the raw coal composition and the X-ray fluorescence analysis for the coal ash composition showed that the raw coal contained 3.84% sulfur and 0.28% Ca.

The pyrolysis experiments were conducted in a tubular fixed-bed reactor. Before and during the pyrolysis, $N_2$ gas was flowed from the bottom to the top of the reactor in a flow rate of 200 ml/min to carry out the pyrolysis experiment under an inert atmosphere. The samples were heated from room temperature to the predetermined temperature in a heating rate of 15 ℃/min. The raw coal, $Ca(OH)_2$-added coal and $CaCO_3$-added coal, were compared in this paper. For simplicity, the $CaCO_3$-added coal is abbreviated to CC coal, and $Ca(OH)_2$-added coal to CH coal. The chars prepared from CC and CH coals were abbreviated to the CC char and the CH char, respectively. Three CC coals were prepared with the weight percentage of $CaCO_3$ of 5%, 10%, and 15%, respectively. Similarly, three CH coals with the weight percentages of $Ca(OH)_2$ of 5%, 10%, and 15% were prepared. The ratios of the inherent Ca content in the raw coal to the added Ca content in CC coals were 1:7.4, 1:15 and 1:25, respectively, while those in CH coals were 1:10, 1:20 and 1:34, respectively.

Most Ca K-edge XANES spectra were collected at Beamline 4B7A of Beijing Synchrotron Radiation Facility (BSRF). Some were collected at the Soft X-ray Micro-characterization Beamline (SXRMB) of Canadian Light Source (CLS). The incident X-ray energies were all calibrated by using white line of calcium sulfate at 4045.3 eV at the two beamlines (BSRF-4B7A and CLS-SXRMB). The collected XANES spectra, respectively, at both beamlines are identical for the same samples. One carboxyl-bound Ca (calcium acetate) and some inorganic minerals ($CaCO_3$, $Ca(OH)_2$, $CaSO_4$, $CaO$ and $CaS$) were selected as the reference compounds. The Ca K-edge XANES spectra of Ca-added chars and the reference compounds were all collected in total electron yield (TEY) mode. After all the XANES spectra were normalized, the XANES spectra of Ca-added chars were fitted with the XANES spectra of reference compounds by using Athena program package [13].

## 3 Results and discussion

The content of Ca in raw coal is only 0.28%. The amounts of $Ca(OH)_2$ and $CaCO_3$ added to raw coal were all controlled to 5%, 10% and 15%. Since the Ca content in raw coal is much small as compared to the loading of $Ca(OH)_2$ and $CaCO_3$ in CH coals and CC coals, the influence of the internal calcium in coal can be ignored. Therefore, the Ca species in the raw coal and in the coal chars prepared from raw coal will not be discussed here.

Fig. 1 shows the Ca K-edge XANES of the CC char prepared at 600 ℃. The fitting results demonstrate that the CC char contains $CaCO_3$, together with calcium acetate, $Ca(OH)_2$, and $CaSO_4$. $CaCO_3$, carboxyl-bound Ca, $Ca(OH)_2$, and $CaSO_4$ accounted for 27±5%, 24±5%, 21±5% and 28±5% of the total Ca in the CC char, respectively. The formation of new species carboxyl-bound Ca, $Ca(OH)_2$, and $CaSO_4$ implies that a majority of $CaCO_3$ in the CC char was decomposed under the condition of coal pyrolysis at 600 ℃. Especially, the formation of carboxyl-bound Ca accompanied by the decomposition of CaCO$_3$ had not been reported before. The formation of the carboxyl-bound Ca is likely due to the reaction of CaCO$_3$ with the carbon matrix of coal during pyrolysis. The conversion mechanism of CaCO$_3$ to carboxyl-bound Ca during pyrolysis of the CC char deserves further investigation. The formation of CaSO$_4$ could be ascribed to the reaction of the additive CaCO$_3$ with the sulfur-containing gases released from the coal during pyrolysis. No CaO was found in the CC char prepared at 600 ℃, because the char sample was exposed to air, leading to hydration of CaO to Ca(OH)$_2$. Similarly, CaS was not detected in this CC char. Probably, CaS was oxidized into CaSO$_4$ by oxide in coal during pyrolysis.

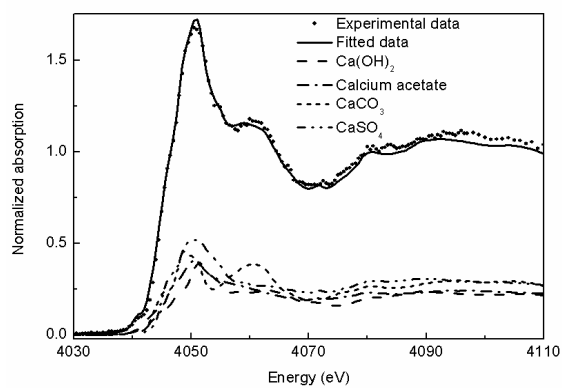

Fig. 1 Ca K-edge XANES of the CC char prepared at 600 ℃

Fig. 2 shows the Ca K-edge XANES of the CH char prepared at 600 ℃. In this CH char, Ca takes the forms of Ca(OH)$_2$, carboxyl-bound Ca, CaCO$_3$, CaO, CaSO$_4$ and CaS. The fitting results indicate that Ca(OH)$_2$ occupies only 18±5% of the total Ca in the CH char, and carboxyl-bound Ca, and CaCO$_3$ accounted for 42±5%, 23±5% of the total Ca in the CH char, respectively. Ca(OH)$_2$ could be also converted to carboxyl-bound Ca. The formation of carboxyl-bound Ca seems to be more effective for CH coal than CC coal. CaSO$_4$ was also formed in the CH char, but the percentages of CaSO$_4$ in the total Ca of the CH char was less than 5%, which was smaller than that in the CC char. The presence of CaO in the CH char is because Ca(OH)$_2$ can be extensively decomposed above 500 ℃, although it might be re-hydrated during the sample storage and use. Some CaS was detected in the CH char, differing from the corresponding char sample derived from CC coal. This suggests that Ca(OH)$_2$ is a more effective sulfur-retention additive. The reaction between Ca(OH)$_2$ and H$_2$S gas released from the coal pyrolysis results in the formation of CaS. This result is consistent with the results of sulfur K-edge XANES in Liu et al. [5]. The formation of CaCO$_3$ in the CH char could be attributed to the reaction between Ca(OH)$_2$ and CO$_2$.

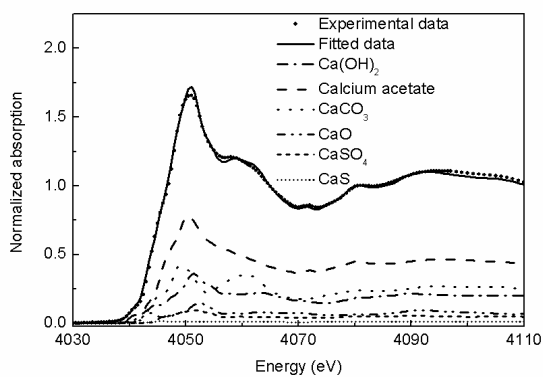

Fig. 2 Ca K-edge XANES of the CH char prepared at 600 ℃

Fig. 3 shows the Ca K-edge XANES of the CH char prepared at 900 ℃. It indicates that Ca in the CH char prepared at 900 ℃ is still in the form of carboxyl-bound Ca, $CaCO_3$, $Ca(OH)_2$, $CaSO_4$, CaO and CaS, which is the same as in the CH char prepared at 600 ℃. It is of great curiosity that the content of carboxyl-bound Ca in the CH chars does not change with the pyrolysis temperature rising from 600 ℃ to 900 ℃. This implies us that calcium is probably dispersed on the surface of the chars through some chemical bonding with oxygenated groups. This type of calcium may be closely related to the catalytic activity of calcium during the coal pyrolysis and char gasification. $CaCO_3$ content in the CH char decreases above 600 ℃ and reduces to less than 10% of the total Ca in the CH char prepared at 900 ℃. The content increase of $Ca(OH)_2$ is consistent with the result of the XRD [5]. $CaSO_4$ is still less than 5%. More CaS is formed at 900 ℃ and its content is up to 12% of the total Ca in the CH char, indicating that the increase of temperature from 600 ℃ to 900 ℃ accelerates the reaction between calcium and $H_2S$. It is reasonable that the content of CaO is little changed.

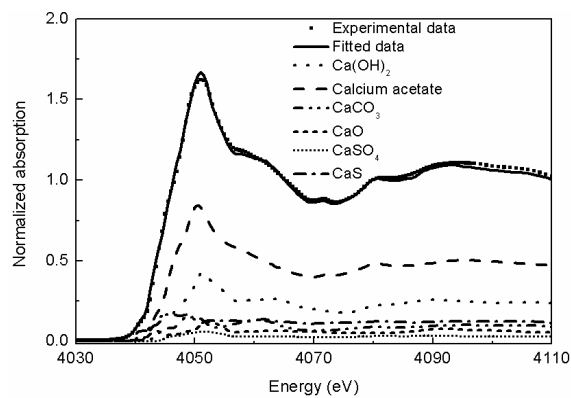

Fig. 3 Ca K-edge XANES of the CH char prepared at 900 ℃

**4 Conclusion**

In this study, Ca K-edge XANES was used to determine the Ca species in coal chars. The results demonstrate that $CaCO_3$ and $Ca(OH)_2$ can be converted to carboxyl-bound Ca during pyrolysis of the Ca-added coal. $CaCO_3$, $Ca(OH)_2$ and $CaSO_4$ always exist in both the CC and CH chars. The content of $CaSO_4$ in the CC char is higher than in the CH char. However, CaS and CaO are only found in the CH chars, indicating that $Ca(OH)_2$ in the CH coal is easily converted to CaO. At the same time, $Ca(OH)_2$ is also easier to capture the $H_2S$ gas to form CaS with the pyrolysis temperature increase of the CH coal.

*The authors wish to thank the staff at Canadian Light Source (CLS) for their assistance, and CFI, NRC, NSERC, OIT, and University of Saskatchewan for the financial support given to CLS.*